%% file: sn-bibliography.tex
\documentclass[pdflatex,sn-mathphys]{sn-jnl}

\usepackage{graphicx}
\usepackage[caption=false]{subfig}
\usepackage{amsmath}
\usepackage{hyperref}
\usepackage{float}

\jyear{2022}%

\theoremstyle{thmstyleone}%
\theoremstyle{thmstyletwo}%

\theoremstyle{thmstylethree}%

\begin{document}

\title[Investigating the Impact of Backward Strategy Learning in a Logic Tutor]{Investigating the Impact of Backward Strategy Learning in a Logic Tutor: Aiding Subgoal Learning towards Improved Problem Solving}


\author*[1]{\fnm{Preya} \sur{Shabrina}}\email{pshabri@ncsu.edu}

\author[1]{\fnm{Behrooz} \sur{Mostafavi}}\email{bzmostaf@ncsu.edu}
\author[1]{\fnm{Mark} \sur{Abdelshiheed}}\email{mnabdel@ncsu.edu}

\author[1]{\fnm{Min} \sur{Chi}}\email{mchi@ncsu.edu}

\author[1]{\fnm{Tiffany} \sur{Barnes}}\email{tmbarnes@ncsu.edu}

\affil*[1]{\orgdiv{Computer Science}, \orgname{North Carolina State University}, \orgaddress{\city{Raleigh}, \postcode{27695}, \state{North Carolina}, \country{USA}}}


\abstract{Learning to derive subgoals reduces the gap between experts and students and makes students prepared for future problem solving. Researchers have explored subgoal labeled instructional materials with explanations in traditional problem solving and within tutoring systems to help novices learn to subgoal. However, only a little research is found on problem-solving strategies in relationship with subgoal learning. Also, these strategies are under-explored within computer-based tutors and learning environments. Backward problem-solving strategy is closely related to the process of subgoaling, where problem solving iteratively refines the goal into a new subgoal to reduce difficulty. In this paper, we explore a training strategy for backward strategy learning within an intelligent logic tutor that teaches logic proof construction. The training session involved backward worked examples (BWE) and problem-solving (BPS) to help students learn backward strategy towards improving their subgoaling and problem-solving skills. To evaluate the training strategy, we analyzed students' 1) experience with and engagement in learning backward strategy,  2) performance, and 3) proof construction approaches in new problems that they solved independently without tutor help after each level of training and in post-test. Our results showed that, when new problems were given to solve without any tutor help, students who were trained with both BWE and BPS outperformed students who received none of the treatment or only BWE during training. Additionally, students trained with both BWE and BPS derived subgoals during proof construction with significantly higher efficiency than the other two groups.}




\maketitle

\section{Introduction}
Attaining the skill to define subgoals is an important component of learning that leads to efficient problem-solving. Defining subgoals or 'subgoaling' in problem-solving refers to the process of decomposing a problem into smaller and easier sub-problems where each sub-problem has a contribution to the overall goal~\citep{catrambone1998subgoal}, or the process of refining the overall goal such that the refined goal (i.e. subgoal) eliminates or reduces the difficulty in achieving the original goal directly~\citep{laird2012universal, newell1994unified, vanlehn1988toward}. Experts produce subgoals during problem-solving more efficiently and easily, and attaining the skill to generate subgoals can induce expert-like behavior in novices~\citep{margulieux2016employing, catrambone1998subgoal}. Thus, researchers from different educational domains (mathematics, statistics, probability, geometry, programming, etc.) have explored varied methods to include subgoaling during problem-solving, such as subgoal labeled examples~\citep{catrambone1998subgoal, catrambone1995aiding}, expert explanations for subgoals, and asking students to write explanations for given subgoals~\citep{margulieux2017using, margulieux2021scaffolding}, etc.). 
These studies suggest that subgoal-infused tutoring methods helped to improve novice performance, using test scores to measure learning. However, these score-based evaluations do not provide enough insight into how these methods impacted students' subgoaling skills.

Additionally, existing research scarcely investigated problem-solving strategies as a way to improve students' subgoal generation skills. Only a few research studies were found that compared expert and novice problem-solving strategies [for example, ~\citep{sweller1988cognitive, heyworth1999procedural}], or investigated subgoal generation strategies in general~\citep{jensen1987stuck}. The two most common problem-solving strategies found in literature are forward and backward chaining. Forward chaining consists of starting from the givens in the problem description and working towards the goal at each step. Backward chaining consists of starting from the goal and refining the goal at each step until the refined goal can be justified by the givens. In other words, backward chaining can be seen as refining the goal to form subgoals at each backward step. Research on human cognitive processes suggests that students try to think backwards more when they need to refine the goal~\citep{sweller1988cognitive, heyworth1999procedural}. However, they find carrying out a backward step to refine the given goal (i.e. subgoaling) difficult during problem-solving ~\citep{trafton1991providing, anderson2014geometry, matsuda2005advanced}. These are the few research we know of that investigates difficulties faced by students when applying backward strategy during problem-solving. 

Unlike prior studies, the main purpose of this study is to explore backward (BW) strategy learning as a medium to aid subgoal learning to improve students' problem-solving skills. To serve this purpose, we designed a training session within an intelligent logic tutor, Deep Thought, where we infused backward worked examples for demonstration of the strategy and backward problem solving for practice. Then, we investigated the impact and efficacy of our training strategy based on: 1) students' experience during the training; 2) students' score-based performance in new problems after training; and 3) student approaches to proof construction. We identify efficient subgoal derivation using a graph mining approach called Approach Maps~\citep{eagle2014exploring}. The key finding of our evaluation suggests that, although backward strategy learning can be difficult for students, it can improve students' subgoaling and problem-solving skills. The main contributions of this study are: 1) an efficient training strategy for backward strategy learning to improve subgoaling and problem-solving skills which can be easily adapted for tutors from other structured problem-solving domains; 2) demonstration of a graph mining based approach analysis that helped us answer how the training session impacted students' subgoaling skill; and 3) important insights on students' problem solving approach obtained from the evaluation results that could be helpful in problem and training session design within automated tutors to improve students' experience and skills.   

\section{Related Work}\label{sec:RW}
\textbf{Importance of Subgoals in Learning and Problem Solving:} Existing literature claims that providing subgoals reduces students' cognitive load and helps them perform better~\citep{margulieux2012subgoal, atkinson2000learning}. Here, cognitive load refers to the load on a students' working memory during learning through problem-solving ~\citep{van2010cognitive}. Thus, researchers from varying domains (including mathematics, natural, and computer sciences) explored subgoal-based instruction during traditional problem solving, and within tutors/learning environments to improve student performance. Margulieux et al.~\citep{margulieux2012subgoal} showed that subgoal-labeled materials reduced students' cognitive load and helped them to be better in programming problem-solving in Android App inventor~\citep{wolber2011app}. Here, a subgoal label refers to a name applied to a set of steps in a problem solution, that segments the overall solution, reducing difficulty~\citep{catrambone1998subgoal}. Margulieux and Catrambone~\citep{margulieux2014improving} also explored subgoal-labeled instructional text coupled with subgoal-labeled worked examples as instructional support within Android App Inventor. They found that students receiving the support outperformed students who do not receive it. Catrambone and Holyoak~\citep{catrambone1990learning}  found evidence that providing subgoals during problem-solving can also be helpful in the transfer of problem-solving skills in domains like algebra and probability. Morrison et al.~\citep{morrison2016subgoals} conducted an experiment where they compared giving subgoal labels against asking students to generate subgoals while they solved programming puzzles, called Parson's problems, that require ordering given pieces of code. They found that students who received subgoal labels performed better than those who had to generate subgoals in low cognitive load Parson's problem post-assessments.  

Zhi et al.~\citep{zhi2018reducing} proposed a data-driven algorithm for subgoal extraction using prior student programming problem attempts. Marwan et al.~\citep{marwan2020adaptive} used this technique to extract subgoals, and presented those along with programming tasks, within a block-based programming environment, iSnap. They found evidence of better performance, higher task completion rate, and less idle time when subgoals were presented in the system. Additionally, Shabrina et al.~\citep{shabrina2020impact} found evidence that when subgoals and subgoal completion-based feedback are presented in iSnap, students closely followed the subgoals, and tried to achieve them, which shaped their approach and interaction with the environment. Cody and Mostafavi~\citep{cody2017investigating} provided subgoals during logic proof construction within an intelligent logic proof tutor. Contrary to research that found positive results with subgoals, they observed that students who received subgoals skipped more problems, and had a significantly higher dropout rate.

\textbf{Aiding Subgoal Learning or Subgoaling:} Although subgoals can reduce excessive cognitive load during problem-solving and thus improve students' performance, they can hinder the learning process and make students unfit for solving future problems~\citep{renkl2002learning}. Thus, researchers also explored methods that might help students to learn subgoaling so that they can form subgoals themselves for a new problem. Existing studies showed that subgoal labels that are not context-specific but are more abstract, are most effective in fueling transfer and helping students to learn subgoaling~\citep{catrambone1996generalizing,catrambone1995aiding}. Richard Catrambone~\citep{catrambone1996generalizing,catrambone1994improving,catrambone1998subgoal} showed that worked examples with abstract subgoal labels for groups of steps helped students to learn subgoaling better and these students were able to successfully transfer the skill to problems that follow a different procedure than what they did during training. Morrison et al.~\citep{morrison2015subgoals} explored two instructional methods in introductory programming tasks: 1) subgoal labels given with the task, and 2) requiring students to generate their own subgoals. Their hypothesis of the first group performing better in posttests was only partially supported by statistical analysis results. In a recent study, Margulieux and Catrambone~\citep{margulieux2017using,margulieux2021scaffolding} showed that students learned better in posttests when they were presented with subgoals and asked to write explanations for the subgoals, 
 when compared to generating their own subgoals.

The main takeaway from existing research focused on helping students to learn subgoaling is that abstract or context-free subgoals aid students better in learning the subgoaling procedure. Also, students learn better when they generate explanations for subgoals themselves rather than when they are given explanations. However, adding the requirement for generating self-explanation during training showed evidence of student struggle as measured in terms of spent time. 

\textbf{Problem-Solving Strategies and Learning to Subgoal:} In a backward strategy or backward chaining, problem-solving is carried out starting from the goal and in each step, the goal is refined to a new subgoal until the initial problem state is reached. Backward strategy is often compared with means-ends analysis~\citep{newell1972human}, which involves carrying out steps to reduce the difference between givens and the goal of a problem. Existing literature claims experts can form subgoals while working in the forward direction due to their high prior knowledge in a domain~\citep{sweller1988cognitive, heyworth1999procedural, larkin1980expert, chi1981categorization}. Also, they may switch between forward and backward strategies during problem-solving ~\citep{sweller1988cognitive}. Prior work has shown that knowing how and when to use each problem-solving strategy is a sign for preparation for future learning~\cite{abdelshiheed2021preparing, abdelshiheed2020metacognition}. However, due to low prior knowledge, novices tend to use backward strategy to figure out substeps of a problem~\citep{heyworth1999procedural}. Matsuda et al. \citep{matsuda2005advanced} explored both forward and backward strategy in a geometry theorem proving tutor and observed that students who learned forward strategy performed better than those who learned backward strategy. They concluded being efficient in using backward strategy is hard for students as they face difficulty in coming up with unjustified statements (subgoals) that are to be proven next.

In this study, unlike prior research, instead of exploring instructional methods, we aimed to aid subgoal learning by infusing a training that induces students to learn backward strategy through demonstration (using backward worked examples (BWE)), and practice (using backward problem solving (BPS)) within Deep Thought. The training phase was long, involving 20 logic proof construction problems that should provide students with enough time to master or adjust to the strategy. We evaluated our training procedure based on students' experience/responses to the training, and test score-based performance. Additionally, unlike existing studies, we investigated efficiency in subgoal derivation while students solved a new problem using approach map analysis.

\section{Method}

\subsection{Deep Thought, The Logic Proof Tutor}
\begin{figure}
     \centering
     \includegraphics[width=\linewidth]{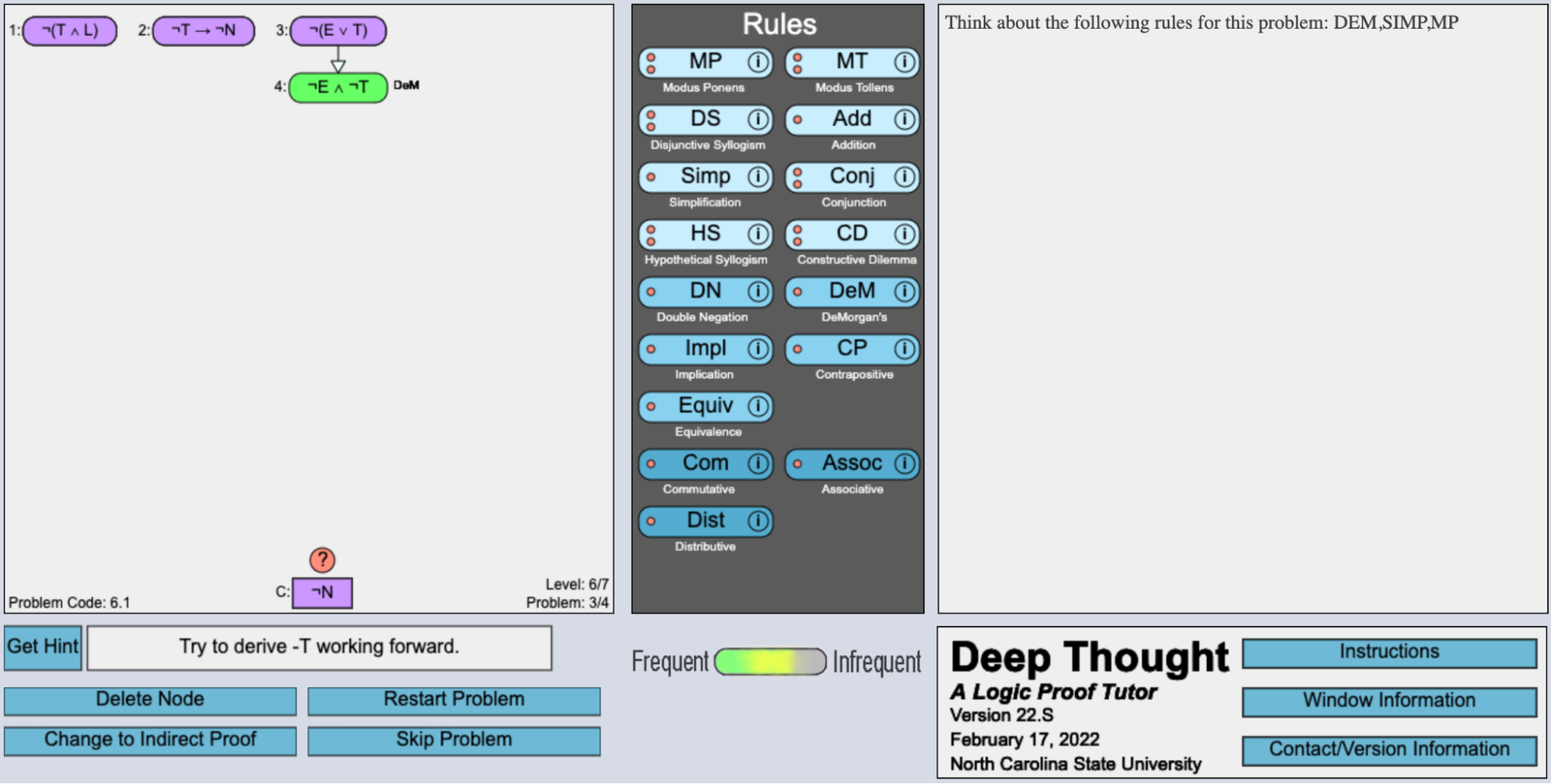}
     \caption{Deep Thought Interface}\label{fig:DT}
\end{figure}

\begin{figure}
    \centering
    \subfloat[WE]{\includegraphics[width=0.42\textwidth]{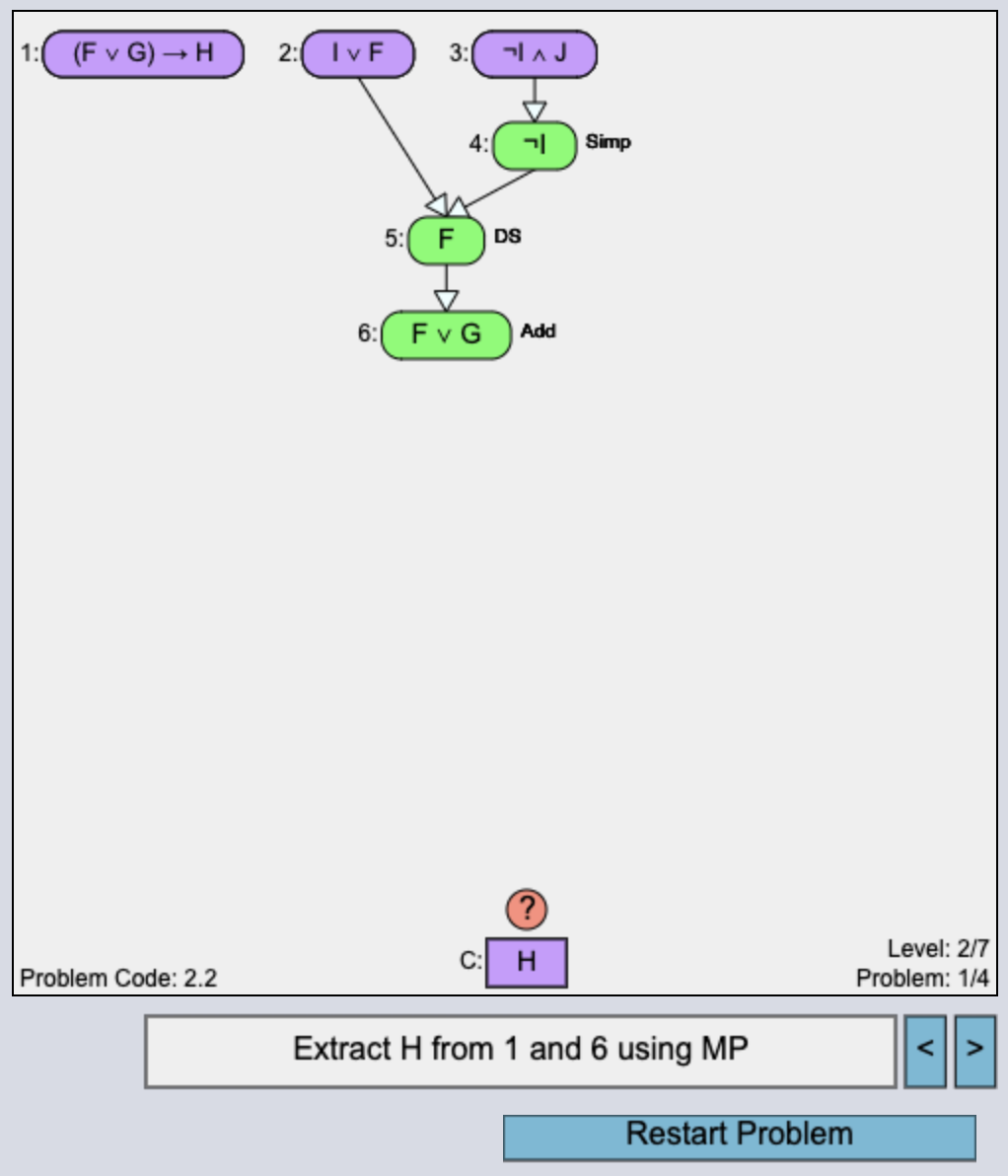}} \hfill
    \subfloat[PS]{\includegraphics[width=0.49\textwidth]{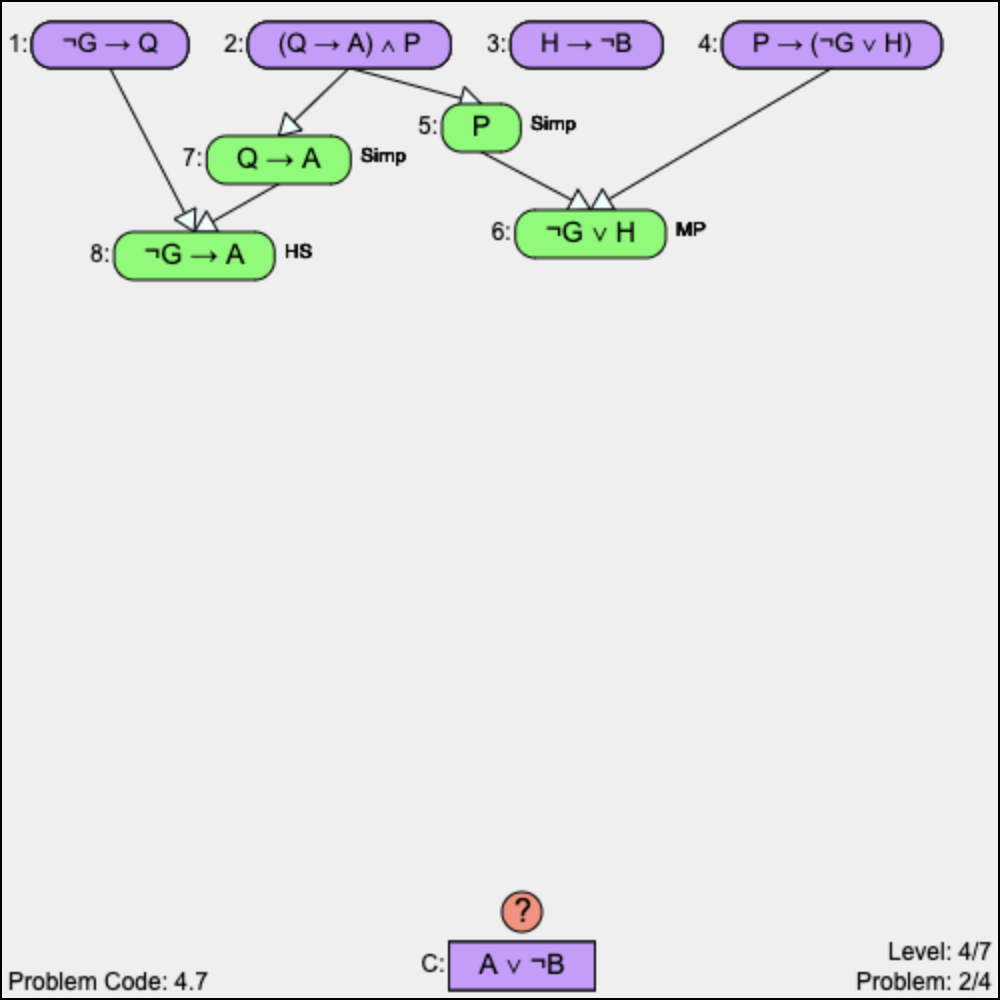}}
    \caption{Proof types within DT: (a) Worked Example (WE); and (b) Problem Solving (PS)}
    \label{fig:WE_PS}
\end{figure}

\begin{figure}
    \centering
    \subfloat[BWE]{\includegraphics[width=0.40\textwidth]{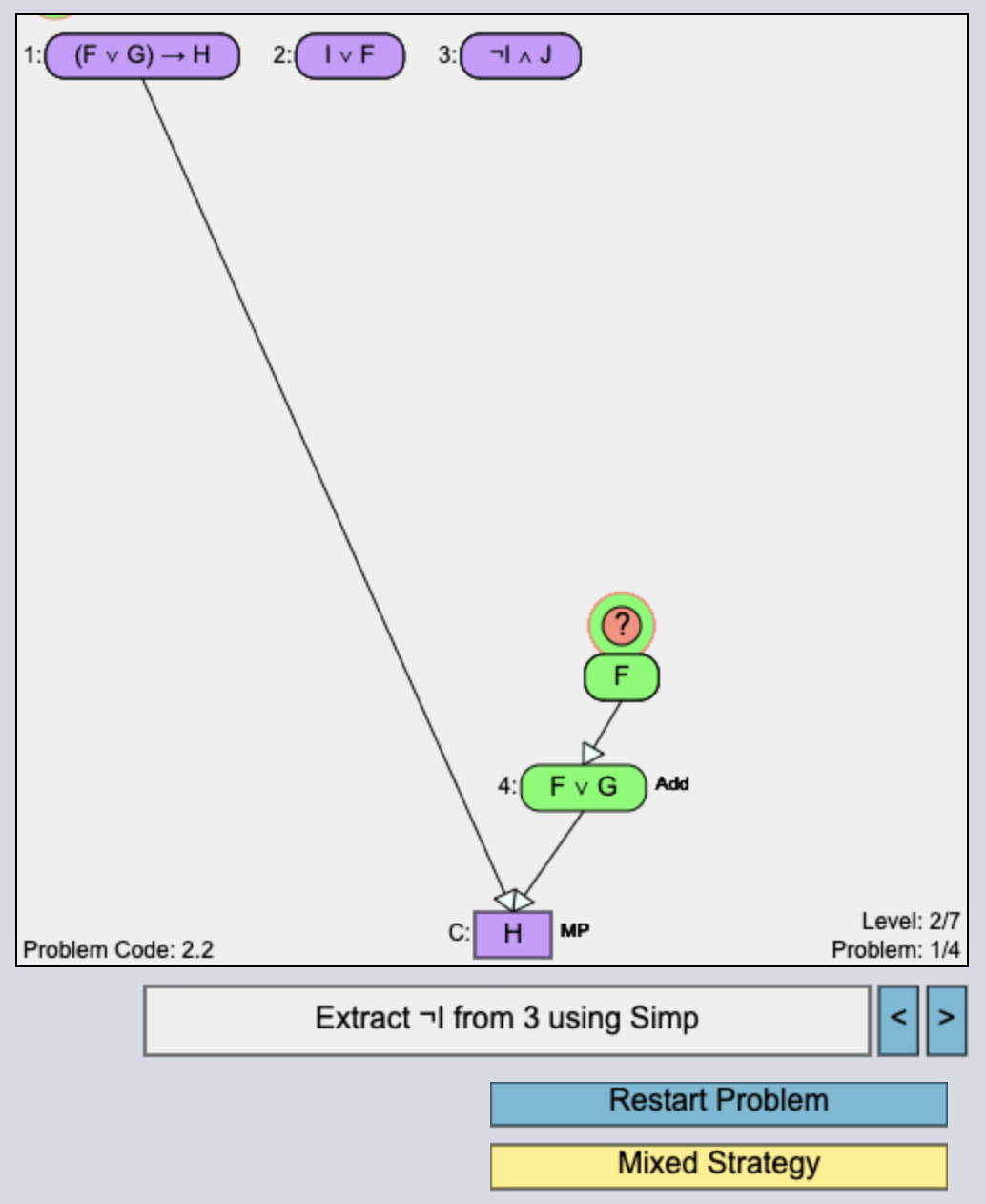}} \hfill
    \subfloat[BW]{\includegraphics[width=0.49\textwidth]{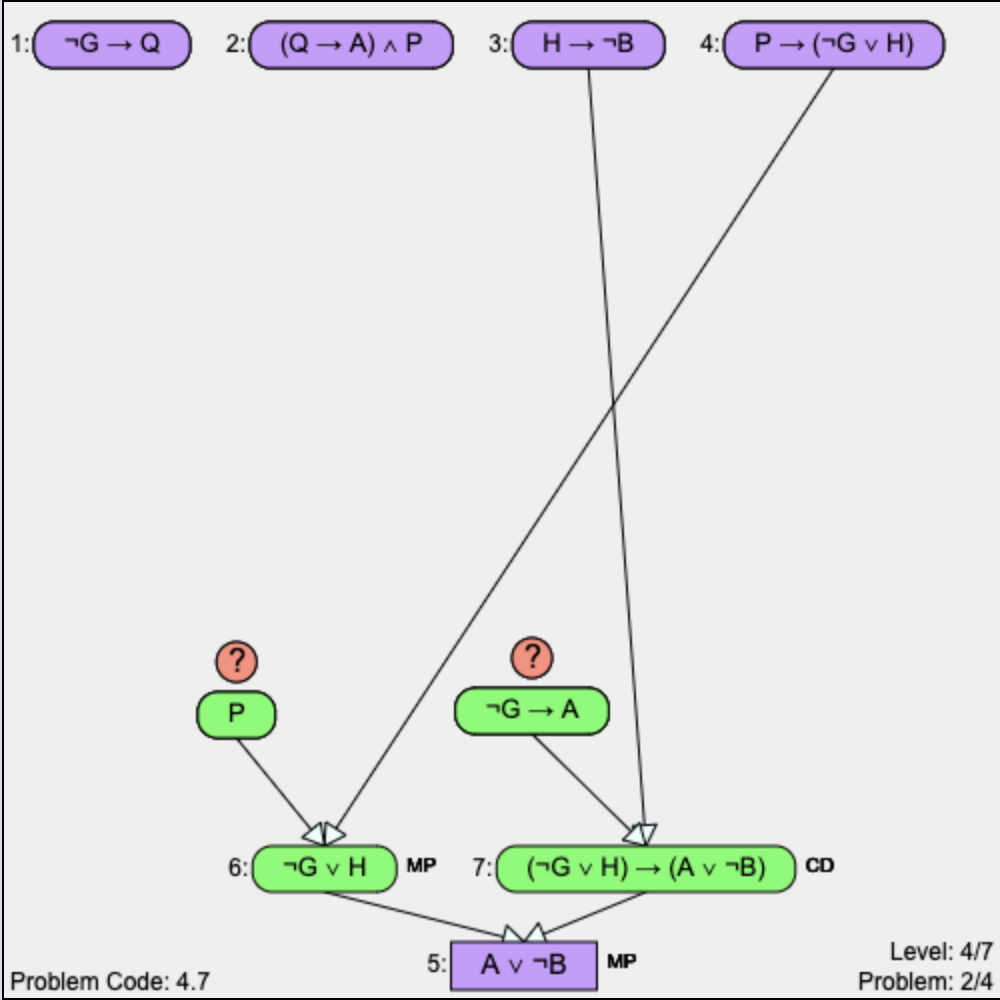}}
    \caption{New Proof types within DT: (a) Backward Worked Example (BWE); and (b) Backward Problem Solving (BPS)}
    \label{fig:BWE_BPS}
\end{figure}

We conducted our study using an intelligent logic tutor, Deep Thought (DT). In DT [Figure \ref{fig:DT}], students are given logic proof construction problems where the premises and the conclusion to be proved are given as visual nodes. A list of logic rules is provided from where students can select rules to apply on premises to derive new ones to reach the conclusion. During the training levels, the tutor also provides on-demand next-step hints and proactive hints, called assertions, that appear when the system predicts that students need help \citep{maniktala2020extending}. DT is organized into 7 levels: one pretest level with 4 problems, 5 training levels with 4 problems in each level, and one post-test level with 6 problems. Each problem is either of type Worked Example (WE) where the tutor constructs the proof [Figure \ref{fig:WE_PS}a] or Problem-Solving (PS) where students are required to construct the proof [Figure \ref{fig:WE_PS}b]. The tutor does not offer any hint or support in the last problem of each level or in the posttest problems. Within DT, logic proofs can be constructed using both forward (FW) [Figure \ref{fig:WE_PS}b] and backward (BW) [Figure \ref{fig:BWE_BPS}b] strategies. 

\begin{figure}
     \centering
     \includegraphics[width=.5\linewidth]{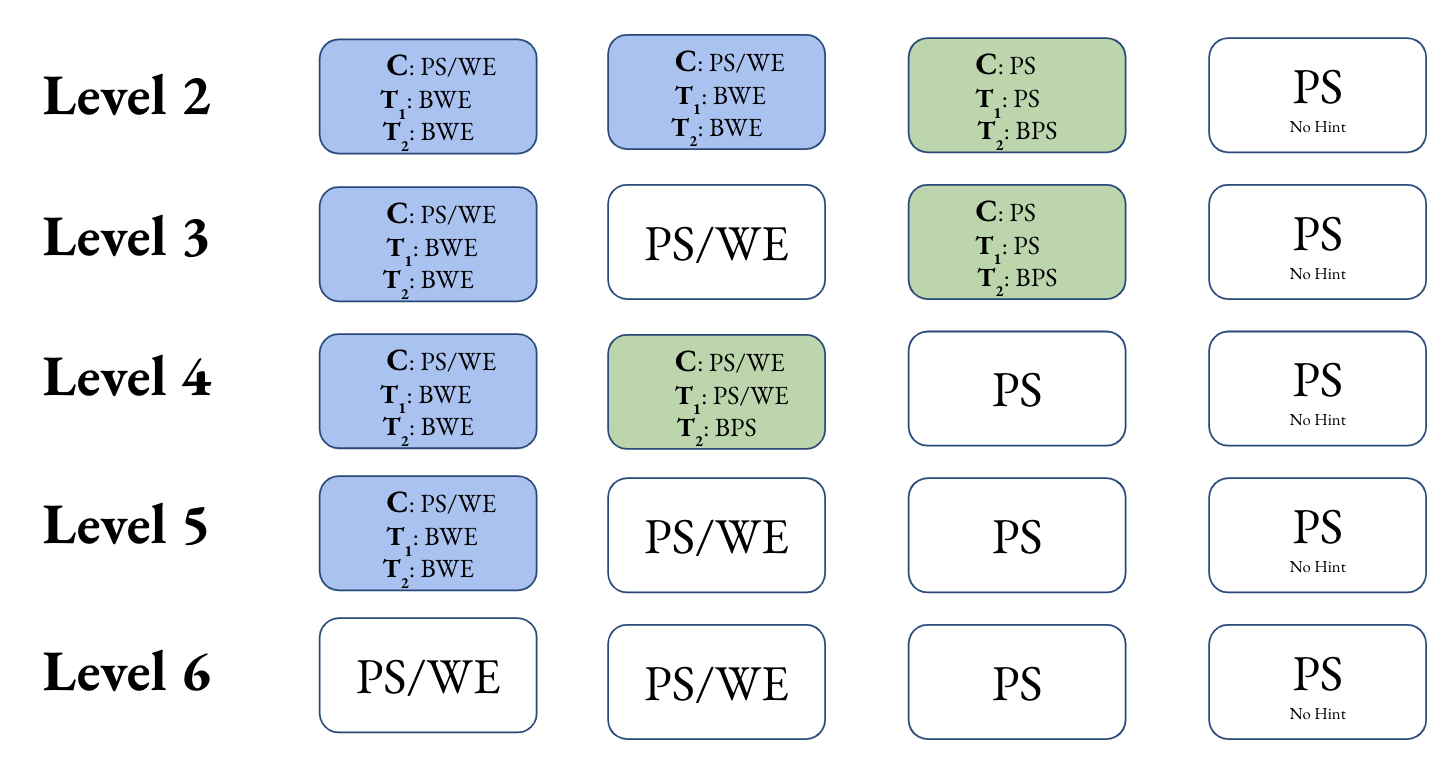}
     \caption{Problem Organization in Training Levels}\label{fig:prob_org}
\end{figure}
 
\subsection{Experiment Design}
For this study, we designed and implemented two new proof types within DT: \textit{Backward Worked Example} (BWE) where the tutor constructs an entire proof using only backward derivation, i.e. the BW strategy, and \textit{Backward Problem Solving} (BPS) where students have to construct the entire proof using only BW strategy. The interface for BWE and BPS are shown in Figure \ref{fig:BWE_BPS}. In these proof types, forward derivations are disabled.

Then, we implemented three training treatments for our experiment: 1) Control (C): this treatment group received only WE and PS problems where they could independently choose to use FW or BW strategy; 2) Treatment 1 - BWE ($T_1$): this treatment group received BWEs in addition to WEs and PSs; and 3) Treatment 2 - BWE+BPS ($T_2$): this treatment group received both BWEs and BPSs along with WEs and PSs. The problem organization in the training levels are shown in Figure \ref{fig:prob_org}. 

\subsection{Deployment and Data Collection}
We deployed DT with our three training treatments in a Discrete Mathematics course for computer science majors offered in a public research university in the United States. The students used DT for a take-home assignment. Each student participating in the course was assigned to one of the three conditions after they completed the pretest level. The assignment algorithm distributes students equally across the treatment groups while ensuring that the pretest scores of the three groups come from the same distribution.

At the end of the experiment, 168 students completed all 7 levels of the tutor with 59 students coming from group C, 55 from $T_1$, and 54 from $T_2$. While students worked in DT, our system collected all information required to replay and reconstruct their proof construction attempts for all problems. The collected data includes all student interactions with the interface (click, selection/deselection of rules/nodes, etc.), and proof steps (derivation/deletion record of nodes with associated predecessors and rules, direction of derivation [FW/BW], spent time, etc.). For each PS problem completed by the students, they were assigned a score that is a function of accuracy and time taken to construct the proof. Note that each logic proof problem given in DT can have multiple solutions where the shortest proof is considered to be the optimal one. Thus, shorter proofs with fewer correct and incorrect rule applications and efficient proof construction in less time received higher problem scores (max score = 100). Note that this score function was devised as a measurement of learning only for research purposes that considers both efficiency in proof construction and optimality of the constructed proofs and has been used in prior research~\citep{abdelshiheed2020metacognition, abdelshiheed2021preparing, sanz2020exploring, cody2018investigation}. However, students' course grades were assigned only based on completion of problems given within DT so that they are not impacted by the experiment.  

\subsection{Research Questions}
The main goal of this study was to aid students' subgoaling skills by introducing and having them learn and practice BW strategy while they construct logic proofs. Our \textbf{hypothesis} was that `Learning BW strategy will improve student performance and make them better prepared for new problem solving by improving their subgoaling skill'. However, prior research~\citep{matsuda2005advanced} states that learning and engaging in BW strategy causes struggle for students. We additionally investigate students' training experience and their response to it. Thus, our investigation on the efficacy and impact of backward strategy learning training focused on the following three research questions:
 \begin{itemize}
\item \textit{\textbf{RQ1 (Students' Experience, and Response):}} How does the backward strategy training impact students' experience in DT, and how do they respond to the training?
\item \textit{\textbf{RQ2 (Impact on Performance):}} How does learning backward strategy impact students' performance in new problems?
\item \textit{\textbf{RQ3 (Impact on Subgoaling Skills):}} How does backward strategy learning impact students’ subgoaling skills?
\end{itemize}
In the subsequent sections, we describe statistical and graphical approach analyses that we conducted to address each of our research questions and corresponding results. Throughout this study, we used Kruskal-Wallis tests to find significant differences ($p<0.05$) across the training groups and performed posthoc pairwise Mann-Whitney U tests with Bonferroni Correction (corrected $p<0.016$) to find an ordering of the groups.\\
\textbf{Note:} To report the results of statistical analyses, we show means as a measure for central tendency, and p-values from pairwise posthoc tests as evidence while comparing the three training groups. For details check out the supplementary materials.

\section{RQ1: Students' Experience, and Response}
\subsection{Training Phase Statistics}
\input{tables/prelim_stat}
To measure students' experience with and response to the training treatments, we calculated step time (avg. time taken to derive one single proposition), problem time, step count, and restart/session counts. We did not find any differences across the groups during pretest. However, during the training phase, we observed significant differences in the metrics when performed Kruskal-Wallis tests with subsequent contrast analyses (pairwise Mann Whitney U tests with Bonferroni correction). Table \ref{tab:prelim_stat} shows the training phase metrics values for PS/BPS problems. Recall that $T_1$, and $T_2$ both were given BWE.  Groups C and $T_1$ received only PS to solve themselves, while $T_2$ additionally received BPS.

We found significant differences in the average of time-related metrics in the training problems. Both step time and overall problem time were significantly higher for $T_2$ than C and $T_1$: in case of problem time, $P_{MW}$($T_2>C$)\footnote[3]{This notation indicates the p-value obtained from Mann-Whitney U test while testing the hypothesis that $T_2$ has significantly higher value than C.}=0.006, and $P_{MW}$($T_2>T_1)<0.0001$; for step time, $P_{MW}$($T_2>C$)=0.004, and $P_{MW}$($T_2>T_1$)=0.0001. This trend in avg. step and problem time of $T_2$ students during training were due to BPS problems [Notice BPS metrics in Table \ref{tab:prelim_stat}]. This implies that carrying out BW steps was difficult for students which required more time during training. Interestingly, $T_1$ and $T_2$ had significantly fewer step counts than C ($P_{MW}<0.0001$, and $P_{MW}$=0.001 respectively for the two cases). These statistics suggests exposure to BW strategy possibly pushed $T_1$ and $T_2$ students towards shorter solution attempts (i.e. thinking/working BW potentially encouraged students to take better steps to reduce the distance between the goal, and given premises).

Additionally, we observed that $T_2$ students required significantly more sessions than C and $T_1$ students ($P_{MW}<0.0001$ in both cases) to complete training problems. Here, each new login or edit separated by a long time period from previous edit is considered a new session. $T_2$ students also had a significantly higher restart count (where students electively start a problem over again) than $T_1$ ($P_{MW}$ = 0.0006) and marginally higher than C ($P_{MW}$=0.02). These statistics suggest $T_2$ students struggled during training due to BPS problems, since BPS problems are restrictive by design, requiring students to construct proofs entirely in backward direction and indirectly encouraging the need for taking the best action at each step.


\subsection{Student Engagement in Backward(BW) Strategy}\label{ss:eng_BW_strat}
\input{tables/bw_stat}
As a measure of students' independent engagement in BW strategy, we used backward (BW) action count which is representative of student intention or attempts to work backwards. We calculated the counts across the three training groups for the $4^{th}$ problem of each training level (Level 2-6), and the 6 post-test problems in Level 7, since no training treatment or tutor help are given in these problems. We report mean BW action counts for each problem across the groups and p-values when significant differences were found in Table \ref{table:bw_engagement}.  

As shown in Table \ref{table:bw_engagement}, $T_2$ students carried out significantly more BW actions than group C and $T_1$ in most of the problems under consideration [2.4, 3.4, 4.4, 5.4, 6.4, and 4 out of the 6 posttest problems: 7.3-7.6]. The group that received backward examples, $T_1$, behaved similarly to C in terms of lower explicit usage of BW strategy. Also, in the earlier phases of training 2.4-4.4, $T_2$ students took too many BW actions ($\sim$31 - $\sim$59 actions) [Table \ref{table:bw_engagement}, column 4; rows 2-4]. However, in the later phases of training, and in the post-test problems, they possibly became calculative in carrying out a BW action as indicated by the reduced count ($\sim$3 - $\sim$11). Our later approach analyses (Section \ref{A_MAP_Analysis}) confirmed that in the early phases of training students used BW strategy inefficiently, with too many BW actions. However, with time $T_2$ possibly adapted to the new skill and was able to use it efficiently (fewer but correct BW steps) to refine goals to subgoals leading to better performance.   

From our analysis of training phase metrics, and students' BW strategy engagement, it is evident that BPS problems posed $T_2$ students a significant amount of struggle (needed more time, more sessions, and more restarts). However, when given new problems, this group voluntarily engaged in BW actions when using the strategy was not even a requirement. On the other hand, group $T_1$ did not seem to face many struggles. But, in terms of BW strategy usage, they behaved similarly to control C. The statistics described above suggest that only BWE may not be motivating, or educational enough for students to attempt to derive propositions in the backward direction. Thus, we conclude that, to successfully motivate students to engage in BW strategy, both examples (BWE), and practice (BPS) are necessary. Additionally, a long training period might be necessary to allow students sufficient time to adapt and become efficient in using the strategy.

\section{RQ2: Impact on Performance}\label{performance_ana}
In this section, we investigate students' performance in relationship with their exposure to BW strategy through BWE ($T_1$), or both BWE, and BPS ($T_2$), or none (no exposure control group C). We calculated students' problem scores across the three training groups over the training and the post-test periods. Again, we focus on the problems where no training treatment/tutor help was given (training PS problems: 2.4-6.4, and post-test problems 7.1-7.6) and students solved them independently with the option of using any of forward (FW) or backward (BW) strategies. Additionally, we calculated problem-solving time, and step count for each problem to investigate the source of higher/lower scores. We report average metric values and significant differences (p-values from Mann-Whitney U tests) across the training groups in Table \ref{table:prob_score} (Problem Scores) and \ref{table:perf_eval} (Problem Time and Step Count).

\input{tables/perf_eval}
\input{tables/prob_score}

\textbf{Problem Scores:} As reported in Table \ref{table:prob_score}, there were no significant differences in problem scores across the groups in the pretest. However, in the earlier phases of training, $T_2$ received significantly lower scores (in prob. 2.4, and 4.4), or lower scores on average (in prob. 3.4) than the other two groups [Table \ref{table:prob_score}, column 2-4;row 2-4]. As the training progressed, $T_2$ outperformed both C, and  $T_1$ (in prob. 5.4), or performed at least as good as them (in prob. 6.4) [Table \ref{table:prob_score}, column 2-4; row 5-6]. In the posttest [Table \ref{table:prob_score}, column 2-4; row 7-12], $T_2$ consistently outperformed the other training groups in all problems (higher avg. in 7.1-7.2, and significantly higher in 7.3-7.6). The problem scores suggest $T_2$ became better at problem-solving over the period of training. On the other hand, $T_1$ mostly performed similar to C (except for 3.4-5.4[insignificant higher avg.]), and did not show consistent signs of improvement.  

Recall that, in problems 2.4-4.4, $T_2$ students were observed to engage in too many BW actions [Table \ref{table:bw_engagement}, column 4;row 2-4]. Note that, the solution of each problem in DT is 5-15 steps long, and too many backward actions indicate unnecessary propositions/actions were explored by the students. However, in the later levels, they engaged in fewer BW actions (possibly only the correct ones) and also received higher scores. This trend suggests that $T_2$ students improved their efficiency over time in using the BW strategy. 

\textbf{Problem Solving Time:} To investigate the source of higher/lower scores, we first analyze problem-solving times across the three training groups. Problem-solving time showed a similar pattern as problem scores. In 2.4 and 4.4, $T_2$ took significantly more time than C and $T_1$ [Table \ref{table:perf_eval}, column 2-4; row 2-4]. This could be possibly due to deriving unnecessary/incorrect steps which were not part of the proof, or due to requiring more time to derive each step. However, as students progressed in DT, problem-solving time for $T_2$ decreased on average (from 5.4-7.2 shown in Table \ref{table:perf_eval}). For problem 7.3-7.6 [Table \ref{table:prob_score}, column 2-4; row 9-12], $T_2$ took significantly lower time than $T_1$ and C while constructing proofs. Due to the similar pattern of problem-solving time and scores, we concluded that learning BW strategy helped $T_2$ students to identify the correct proof construction approach in less time, improving their scores.

\textbf{Step Count:} Logic proof problems within DT can have multiple solutions with different lengths. However, possibly due to having the same level of prior knowledge (measured during pretest), for most problems students were observed to construct similar proofs with similar lengths (details are discussed in Section \ref{A_MAP_Analysis} using Approach Maps). As shown in Table \ref{table:perf_eval} column 6-9, in most of the training, and post-test problems, no significant differences were found in step counts across the three groups. However, we observed, in problems 7.3, and 7.5, $T_2$ had significantly fewer steps than C and $T_1$ [Table \ref{table:perf_eval}, column 6-9; row 9 \& 11]. Note that step count can be different due to the adoption of different solution approaches, or due to different unnecessary/incorrect proposition derivation counts. Our later approach analysis showed that to solve problem 7.3, the shortest student solution was 8 steps long, and $81\%$ of $T_2$ students (44 out of 54 students) adopted the shortest 8-step approach, whereas the percentages for C, and $T_1$ were only $54\%$ (32 out of 59 students), and $49\%$ (27 out of 55 students) respectively. A similar pattern was observed for problem 7.5 where $65\%$ $T_2$ students adopted the shortest 6-step solution, whereas the percentages for C, and $T_1$ were only $35\%$, and $41\%$ respectively. These results suggest that BW strategy has the potential to drive students toward shorter solutions. However, this trait could be dependent on the specific problem a student is working on, since this pattern was only observed in two post-test problems.

The results of our score-based performance analysis suggest that the combination of BWE and BPS improved students' problem-solving skills, and helped them to perform better in the post-test. $T_2$ students obtained higher scores by constructing post-test proofs faster, or by constructing shorter post-test proofs. On the other hand, $T_1$ students who only received BWE, behaved and performed mostly like the control group C, who were not introduced to the backward strategy at all. In the next section, we investigate in more detail student solution approaches to identify the source of students' improved problem-solving skills as demonstrated by test scores.

\section{RQ3: Impact on Subgoal Learning}
\subsection{Approach Map Analysis}\label{A_MAP_Analysis} To investigate $T_2$ students' higher performance (as reported in Section \ref{performance_ana}) in relationship with subgoaling, we generated graphical representations of students' proof construction attempts using \textit{Approach Maps}~\citep{eagle2014exploring}. We identified expert subgoals in those solution approaches and analyzed each proposition derived using statistical tests to identify the instances where one training group was more efficient than another. From our analyses of BW action counts, and performance, we identified four scenarios: \textbf{1)} Poor performance of $T_2$ co-occurring with a lot of BW actions (2.4-4.4); \textbf{2)} Better performance of $T_2$ with no significant differences in BW actions (Prob. 5.4); \textbf{3)} Significant higher performance of $T_2$ co-occurring with comparatively more BW actions due to less problem solving time (7.3-7.6), or \textbf{4)} due to fewer step counts (7.3, and 7.5). In the subsequent subsections, we first describe the construction method of approach maps and then, analyze approach maps of representative problems for each of the scenarios mentioned above.
 
\subsubsection{\textbf{Approach Map Generation Method}}
An approach map, proposed by Eagle et al.~\citep{eagle2014exploring}, is a graphical representation of students' problem-solving approaches. The steps to construct approach maps are briefly described below:

\textbf{Step 1 (Construct Interaction Networks from Students' Action Logs):}
An interaction network~\citep{eagle2012interaction} for a problem is essentially a graph consisting of nodes and edges representing all students' problem-solving states and actions. For DT problems, a state is the set of all propositions (both justified, and backward derived unjustified ones) a student has at any point of the construction of a proof, and an action is the addition or deletion of a proposition through the application of a logic rule. Note that the propositions in a state are lexicographically ordered and the order of their derivation is ignored, since considering the order of derivation could increase the number of states exponentially, and no complex computation would be feasible on the interaction network. 

As students progress in the construction of a proof, they move from state to state via actions. For example, if at state, $S_0( \neg(K \land M), J \Rightarrow (K \land L), L \Rightarrow M)$, the action DeMorgan's rule on $\neg(K \land M)$ is applied, the new state will be $S_1(\neg(K \land M), \neg K \lor \neg M, J\Rightarrow(K \land L), L \Rightarrow M)$. An incorrect rule application can result in the previous state and next state to be the same. The tuple (current state, action, next state) is called an interaction. So, a students’ solution attempt for a logic proof problem is a directed graph of interactions. Our code implementation generates interaction networks for a problem by conjoining all the interactions from student attempts. There is a single start node in the interaction network containing the given premises. There can be multiple end states (since there can be multiple solutions for the same problem) where each end state contains the justified goal statement. Additionally, to facilitate statistical analyses on the network, the interaction network includes data on frequency of node and edge visits, time spent on/before each interaction, and step counts before each interaction across the three training groups.

\textbf{Step 2 (Girvan-Newman Clustering Algorithm):} Students, due to not having expert-like prior knowledge/skills, require exploration leading to derivation/deletion of unnecessary/incorrect propositions along with correct propositions throughout a proof construction attempt. These derivations/deletions form visible clusters/regions in the interaction network where the major outcome of these regions are the proposition(s) contributing to the final proof. Also, different approaches to solving the same problem can result in different regions. To identify these regions, the approach map technique applies the Girvan-Newman community clustering algorithm \citep{girvan2002community} on interaction networks. The clustering algorithm takes as input an interaction network with start/end nodes, and self-loops (edges originating and ending at the same node) removed. Additionally, edge weights are assigned based on the cumulative visit frequency of the corresponding interaction. At each iteration of the algorithm, the edge with the highest edge-betweenness is removed from the network. Edge betweenness~\citep{cuzzocrea2012edge} of an edge is calculated by calculating the shortest paths between all pairs of nodes and counting the number of shortest paths that go through that edge. Then, the connectivity of the resulting graph is calculated using modularity score~\citep{newman2010networks}. Each connected component of the resulting graph is marked as a region. This process is continued till there is no edge left to be removed. The output of the algorithm is the graph with the highest modularity score and the clusters/regions are the connected components within that graph.

\textbf{Step 3 (Approach Map Generation from Clustered Network):} The clustered interaction network for any logic proof problem is a fairly large graph where student approaches to solve the problem cannot be visually detected. Thus, we simplified the clustered networks to approach maps using the method adopted by Eagle et al.~\citep{eagle2012interaction}. First, we added the start and end nodes back to the clustered interaction network and then applied the following steps: \textbf{1)}    Represent each region with a single node and label them with the proposition(s) with the highest number of incoming and outgoing edges from and to other regions, and all propositions derived to generate the latter from the former. This step filters out unnecessary propositions derived by the students, and keeps only the ones contributing to the proof; \textbf{2)} Combine parallel edges, and actions between regions to a single edge with a composite action label; and \textbf{3)} Keep only unique paths between the start and goal nodes. These three steps convert a clustered interaction network to a pseudo-graph called an approach map, where the start node is connected to the goal node via region nodes. A path from the start node to a goal node represents a solution approach, where the propositions contributing to the solution can be visually identified from the labels of the regional nodes in between. 

\begin{figure}
    \centering
    \subfloat[Prob. 2.4]{\includegraphics[width=0.48\textwidth]{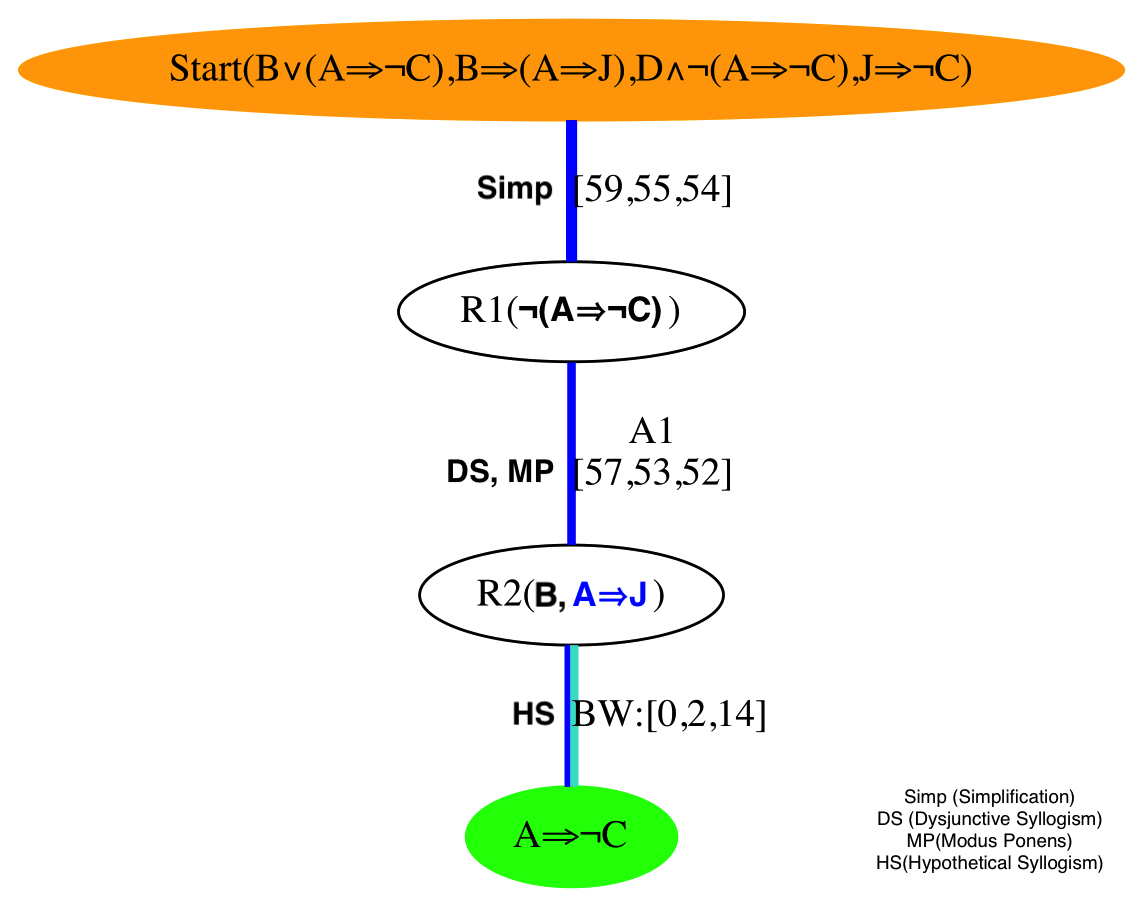}}\\
    \subfloat[Prob. 5.4]{\includegraphics[width=\textwidth]{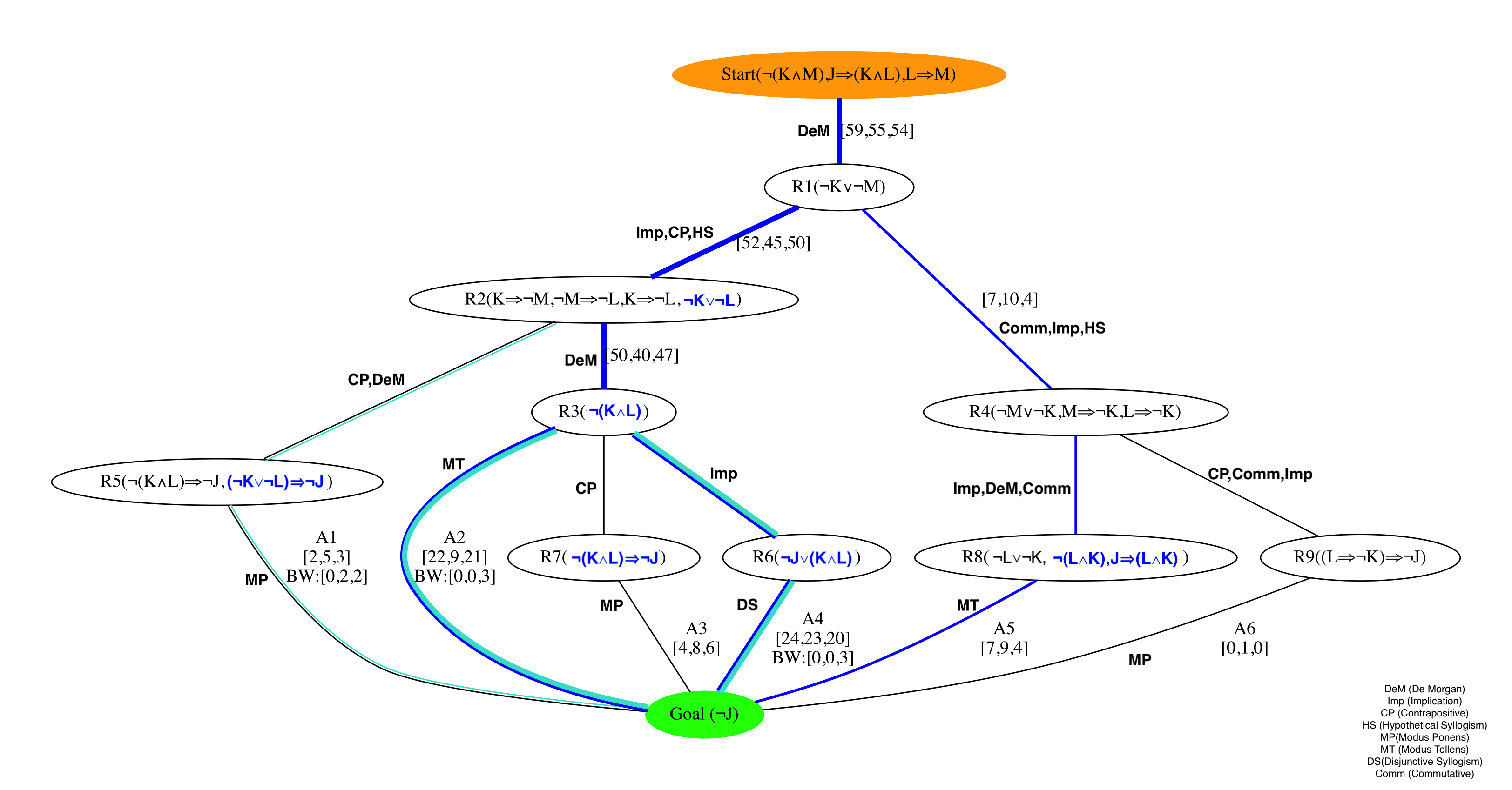}}
    \caption{Approach Maps for Problem a) Problem 2.4, and (b) Problem 5.4}
    \label{fig:approach_maps_train}
\end{figure}

\textbf{Approach Map Presentation:} In the approach maps presented in this paper [for ex. refer to Figure \ref{fig:approach_maps_train}], we only showed the most common student solution approaches and used them to discuss differences found across the training groups. The start node contains the given premises, and the goal node contains the statement to justify. Region nodes are labeled as R1, R2, etc. Each path from Start to Goal represents an approach (labeled as A1, A2, etc.). Each edge is labeled with the applied rule(s), and visit frequency across the training groups as [n(C), n($T_1$), n($T_2$)]. Edge thickness and color are based on visit frequency (frequent edges are thicker and colored blue; non-frequent edges are colored black and of unit thickness). Expert-identified subgoal propositions are colored blue. Bold-faced propositions (Blue/Black) indicate significant differences in derivation across groups. Regions attached to the multicolor edge(s) indicate the last propositions of those regions were sometimes derived backwards by students. BW derivation counts are also attached in such cases. Next, we present approach maps for representative problems discussing different scenarios. 

\subsection{Scenario 1 (Poor Performance of $T_2$ Co-occurring with Many BW actions): Prob. 2.4}
Problem 2.4 asks to derive $A\Rightarrow\neg C$ from $B\lor(A\Rightarrow\neg C), B\Rightarrow(A\Rightarrow J), D\land\neg(A\Rightarrow\neg C), and J\Rightarrow\neg C$. In the approach map of this problem [Figure \ref{fig:approach_maps_train}a], we show the most common three-step solution approach for this problem (adopted by $~96\%$ of the students) labelled as A1 [Start $\rightarrow$ R1($\neg(A\Rightarrow\neg C)$) $\rightarrow$ R2 ($B, A\Rightarrow J$) $\rightarrow$ Goal] in the figure. Note that in this approach, $A\Rightarrow J$ was identified as a subgoal by experts [marked blue in Figure \ref{fig:approach_maps_train}a].

As we analyzed each of the steps in the solution approach statistically, we observed, when working in forward direction, $T_2$ spent significantly more time than C, and $T_1$ to derive each of the three propositions: $\mathbf{\neg(A\Rightarrow\neg C)}$ [mean(C, $T_1$, $T_2$)= 4.98, 5.77, and 9.36 mins.; $P_{MW}$($T_2>C$)=0.0005; $P_{MW}$($T_2>T_1$)=0.0008], \textbf{B} [mean(C, $T_1$, $T_2$)= 4.33, 4.37, and 9 mins.; $P_{MW}$($T_2>C$)=0.003; $P_{MW}$($T_2>T_1$)=0.011], and $\mathbf{A\Rightarrow J}$ [mean(C, $T_1$, $T_2$)= 6.98, 5.77, and 12.36 mins.; $P_{MW}$($T_2>C$)=0.006; $P_{MW}$($T_2>T_1$)=0.015]. $T_2$ students also derived more unnecessary propositions before finding the right direction, and deriving the correct steps [before deriving $\neg(A\Rightarrow\neg C)$, mean unnecessary proposition count (C, $T_1$, $T_2$)= $\sim$1,$\sim$1, and $\sim$4; $P_{MW}$($T_2$>C)=0.001; $P_{MW}$($T_2$>$T_1$)=0.002]; and before deriving B, mean (C, $T_2$)= $\sim$4, and $\sim$6; $P_{MW}$($T_2>C$)=0.009]. However, when $T_2$ students successfully refined the final goal to a subgoal by backward deriving $A\Rightarrow J$ with the given premise $J\Rightarrow\neg C$ using Hypothetical Syllogism (14 instances found of such derivation), their solution attempt had fewer overall unnecessary steps than C, and $T_1$ [mean(C, $T_1$, $T_2$)= 7, 6, and 4; $P_{MW}$($T_2<C$)=0.0005; $P_{MW}$($T_2<T_1$)=0.011]. However, they still needed more time to construct the proof backwards than what C and $T_1$ groups took to derive the proof in forward direction [mean BW derivation time for $T_2$ = 38.9 mins.; $P_{MW}$($T_2>C$)=0.001; $P_{MW}$($T_2>T_1$)=0.006]. Note that no significant differences were found between the solution attempts of $T_1$ and C. 

The statistics above, together with the BW action count presented in Section \ref{ss:eng_BW_strat}, suggest that $T_2$ students attempted to define subgoals using BW strategy explicitly just after the first level of training. However, only a few $T_2$ students (14 students) were successful in deriving the correct subgoal ($A\Rightarrow J$). In this phase, they were still adapting to the BW skill and struggled (needed more time) to identify, and derive correct propositions in the backward direction. In the instances of failed BW derivation attempts, students were observed to derive more unnecessary propositions (in both FW/BW direction) that increased time, and step count decreasing their scores. On the other hand, when students were successful in deriving BW steps, the unnecessary proposition count decreased, but they still needed more time.

\subsection{Scenario 2 (Improved performance of $T_2$ with no Significant Differences in BW actions): Prob. 5.4}
Problem 5.4 asks to derive $\neg J$ from the premises: $\neg(K \land M),J\Rightarrow(K\land L), and L\Rightarrow M$. From the approach map for this problem [Figure \ref{fig:approach_maps_train}b], we identified 6 solution (labelled A1 - A6) approaches. Among these approaches, A2 [Start $\rightarrow$ R1 $\rightarrow$ R2 $\rightarrow$ R3 $\rightarrow$ Goal], and A4 [Start $\rightarrow$ R1 $\rightarrow$ R2 $\rightarrow$ R3 $\rightarrow$ R6] are the most common approaches [covering $~71\%$ of students]. In these two approaches, the common expert subgoal is to derive $\neg(K\land L)$ using Modus Ponens on $\neg K\lor\neg L$. Our statistical analysis showed that group $T_2$ derived both of these propositions in significantly less steps than group C [$\mathbf{\neg K\lor\neg L}$: mean(C, $T_2$)= 9.16, and 7.42; $P_{MW}$($T_2<C$)=0.011; $\mathbf{\neg(K\land L)}$: mean(C, $T_2$)= 10.44, and 8.17; $P_{MW}$($T_2<C$)=0.003]. This suggests group C required more exploration through unnecessary steps than $T_2$ before figuring out what to derive to achieve the final goal. The third most frequent approach in the approach map is A5 [Start $\rightarrow$ R1 $\rightarrow$ R4 $\rightarrow$ R8 $\rightarrow$ Goal] which is lengthier than A2 and A4. In the final step of this approach, Modus Tollens is applied on $\neg(L\land K)$, and $J\Rightarrow(L\land K)$ to derive the final goal ($\neg J$) which makes $\neg(L\land K)$ a subgoal in this approach. Group $T_2$ outperformed both C, and $T_1$ in deriving $\neg(L\land K)$ in terms of time [mean(C, $T_1$, $T_2$)= 26.07, 20.97, and 15.48 mins.; $P_{MW}$($T_2<C$)=0.002; $P_{MW}$($T_2<T_1$)=0.014].

Only 8 $T_2$ students (as shown by the multicolor edges in A1, A2, and A4 [Figure \ref{fig:approach_maps_train}b]) defined subgoals (($\neg K\lor\neg L, \neg K\lor\neg L\Rightarrow\neg J$) in A1, or $\neg(K\land L)$ in A2, or ($\neg(K\land L), \neg J\lor(K\land L)$) in A4) using BW derivation. $T_2$ students who constructed the proof by defining subgoals in this fashion, outperformed C in terms of both step count, and time [\textbf{Step Count:} mean(C, $T_2$)=11.2,and 8.5; $P_{MW}$($T_2<C$)=0.0001; \textbf{Time:} mean(C, $T_2$)= 31.1, and 25.85 mins.; $P_{MW}$($T_2<C$)= 0.008].

These results suggest, that although in this problem, most $T_2$ students did not explicitly derive BW steps, they were efficient in deriving propositions identified as subgoals, possibly through BW thinking which helped to outline the entire proof in less time with fewer unnecessary steps. Also, the few students who explicitly derived BW steps were more successful (fewer unnecessary propositions, and less time) than they were in the previous problem where they needed more time to work backwards (in prob. 2.4). We concluded that at this phase (5th level of training) $T_2$ students who received both BWE, and BPS were better adapted to using BW strategy (explicitly/implicitly) for subgoaling. However, in this problem, although $T_1$ received higher avg. scores than C [Table \ref{table:perf_eval}], we found no significant evidence supporting improvement of group $T_1$.

\begin{figure}
    \centering
    \subfloat[Prob. 7.3]{\includegraphics[width=0.48\textwidth]{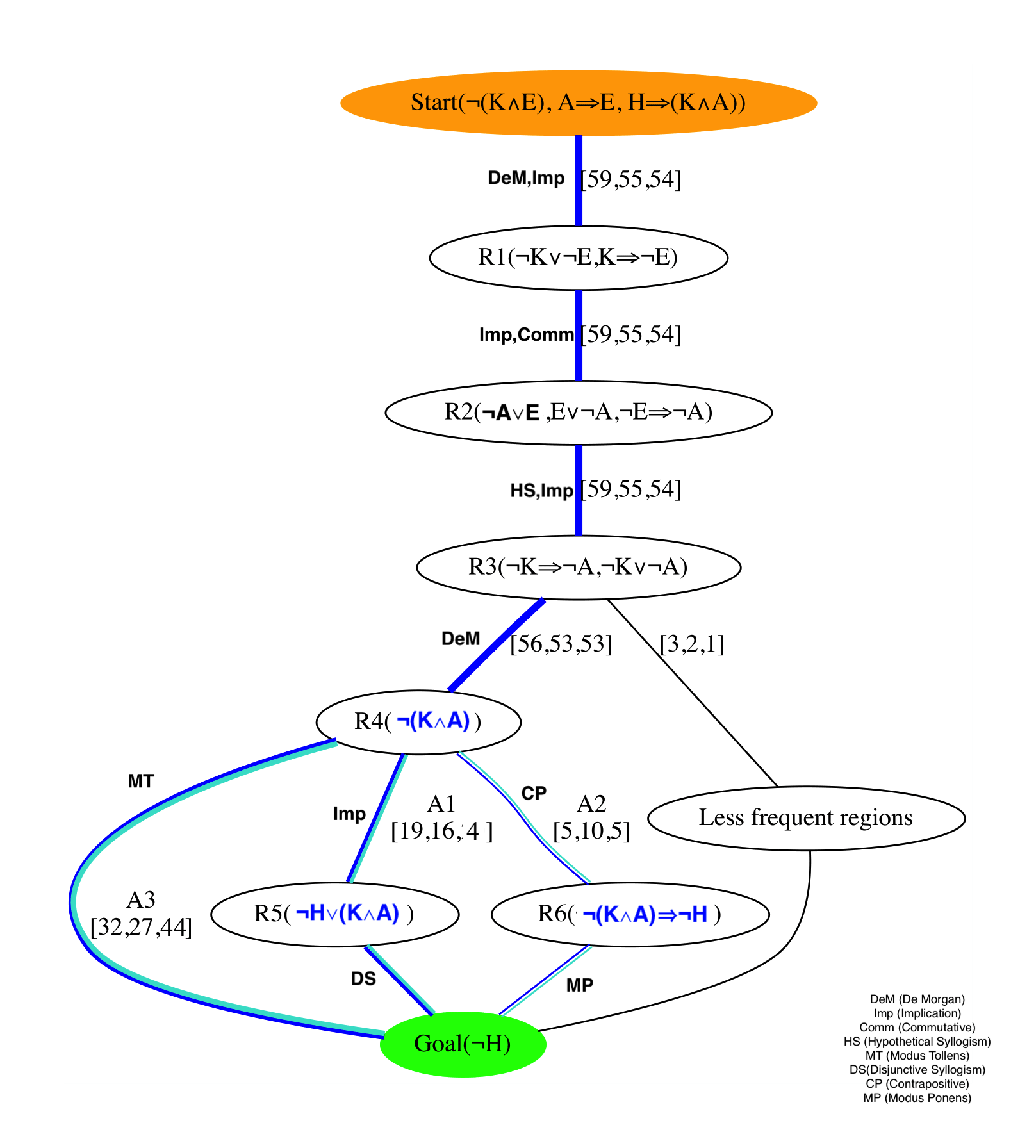}}
    \subfloat[Prob. 7.6]{\includegraphics[width=0.48\textwidth]{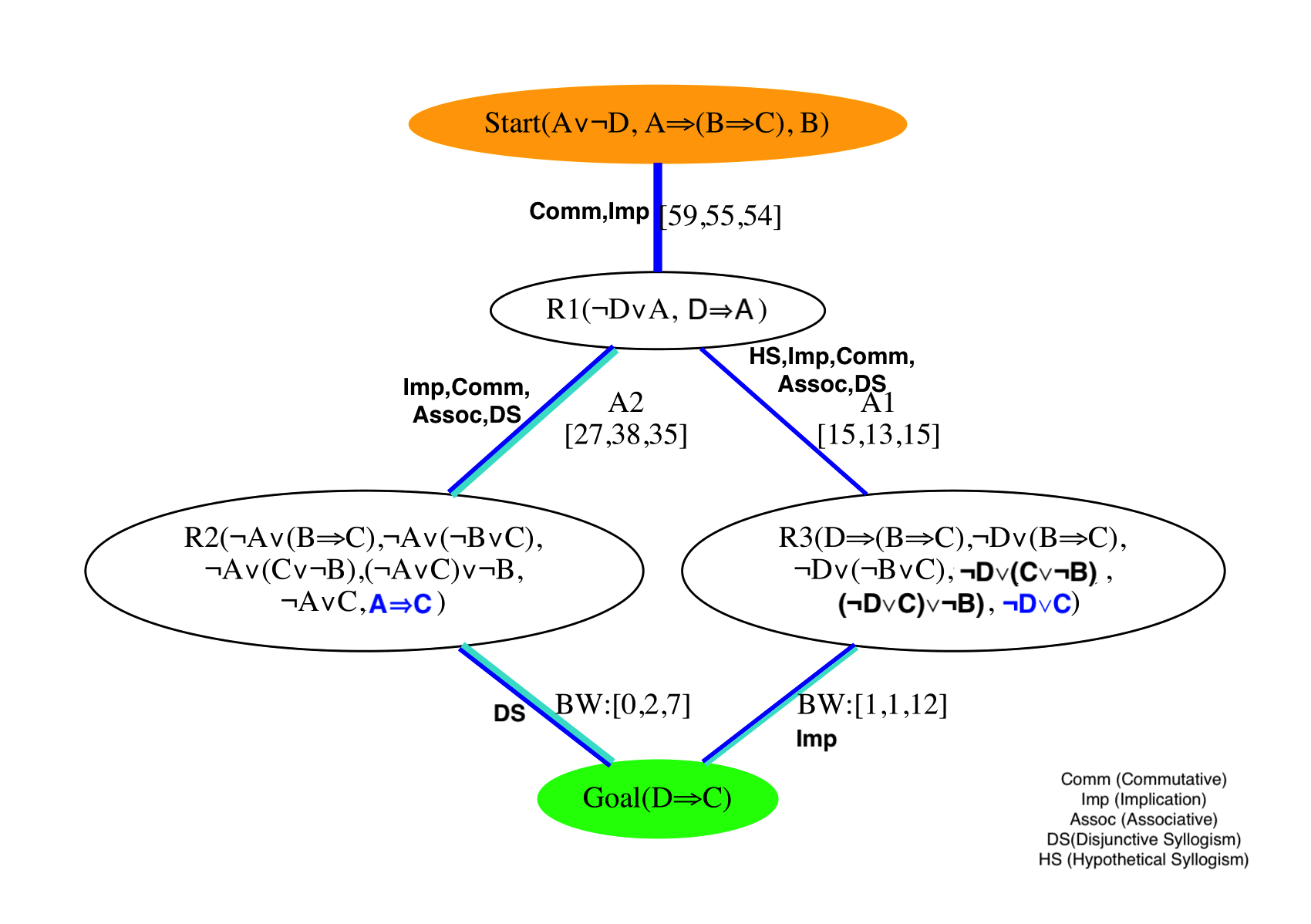}}
    \caption{Approach maps for (a) Problem 7.3; and (b) Problem 7.6}
    \label{fig:approach_maps_posttest}
\end{figure}

\subsection{Scenario 3(Improved Performance of $T_2$ (less time and steps) with Fewer Effective BW Steps): Prob. 7.3}
Problem 7.3 asks to derive $\neg H$ from the premises : $\neg(K\land E), A\Rightarrow E, and H\Rightarrow(K\land A)$. The approach map for this problem [Figure \ref{fig:approach_maps_train}a] shows the three most commonly adopted solution approaches (covers $~96\%$ of all students) labelled as A1 [Start $\rightarrow$ R1 $\rightarrow$ R2 $\rightarrow$ R3 $\rightarrow$ R4 $\rightarrow$ Goal], A2 [Start $\rightarrow$ R1 $\rightarrow$ R2 $\rightarrow$ R3 $\rightarrow$ R4 $\rightarrow$ R5 $\rightarrow$ Goal], and A3 [Start $\rightarrow$ R1 $\rightarrow$ R2 $\rightarrow$ R3 $\rightarrow$ R4 $\rightarrow$ R6 $\rightarrow$ Goal]. One common expert-identified subgoal in these three approaches is $\neg(K\land A)$ or $\neg(A\land K)$ (from region R4). And, statistically, $T_2$ derived $\neg(K\land A)/\neg(A\land K)$ with significantly fewer unnecessary steps than group C and $T_1$ [\textbf{For} $\mathbf{\neg(K\land A)}$, mean(C, $T_1$, $T_2$)= 9.45, 9.46, and 7.38; $P_{MW}$($T_2<C$)=7.91e-05; $P_{MW}$($T_2<T_1$)=0.007; \textbf{For} $\mathbf{\neg(A\land K)}$, mean(C, $T_2$)= 9.19, and 7.64; $P_{MW}$($T_2<C$)=0.01662]. Moreover,  $T_2$ derived $\neg A\lor E$ significantly earlier in their solution attempt than C [mean(C, $T_2$)= 9.60, and 4.53 mins.; $P_{MW}$($T_2<C$)=0.009]. From the approach map, we can see that $\neg(K\land A)$ comes from $\neg K\Rightarrow\neg A$ which is derived by applying Hypothetical Syllogism on $K\Rightarrow\neg E$ and $\neg E\Rightarrow\neg A$. And, early derivation of $\neg A\lor E$ suggests that $T_2$ students also figured out the approach to derive $\neg(K\land A)$ quicker than other groups. 

Additionally, 22 $T_2$ students ($\sim41\%$ of $T_2$ students), explicitly backward derived $\neg(K\land A)$ defining it as a subgoal using Modus Ponens (with $\neg(K\land A)>\neg H$), Modus Tollens (with $H\Rightarrow(K\land A)$ given), or Disjunctive Syllogism (with $\neg H\lor(K\land A)$). Overall, these $T_2$ students had significantly less step count and spent less time than avg. of group $T_1$ and C [\textbf{step count:} mean(C, $T_1$, $T_2$)= 10.5, 9.78, and 7.62; $P_{MW}$($T_2<C$)=1.28e-05; $P_{MW}$($T_2<T_1$)=0.0006; \textbf{problem time:} mean($T_1$, $T_2$)= 25.5, and 15.6 mins.; $P_{MW}$($T_2<T_1$)=0.009]. Also, notice that most $T_2$ students adopted the shortest solution A3 (as mentioned before in section \ref{performance_ana}) while solving this problem. 

In this problem, $T_2$ students engaged in BW derivations comparatively more than C and $T_1$, and they did so efficiently. They also continued to show improved subgoaling behavior. In addition to fewer unnecessary steps and less time, in this problem, we observed $T_2$ students be more driven toward the shortest solution.

\subsection{Scenario 4(Improved Performance of $T_2$ (less time) with Fewer Effective BW Steps): Prob. 7.6} 

Problem 7.6  asks to derive $D\Rightarrow C$ from $A\lor\neg D, A\Rightarrow(B\Rightarrow C)$, and B. From the approach map for this problem [Figure \ref{fig:approach_maps_posttest}b], we identified the two most commonly adopted solution approaches (adopted by $\sim79\%$ of all students), labelled as A1 [Start $\rightarrow$ R1 $\rightarrow$ R3 $\rightarrow$ Goal], and A2 [Start $\rightarrow$ R1 $\rightarrow$ R2 $\rightarrow$ Goal] in the figure.  The common expert-identified subgoal in both A1, and A2 is $D\Rightarrow A$ (from region R1), which is only two steps away from the start point, and we did not find any differences in derivation of this subgoal across our training groups. The other expert subgoal for A1 is $\neg D\lor C$ (from region R3). Twelve (12) $T_2$ students derived this subgoal in BW direction. We observed that group $T_2$ derived this subgoal with significantly less steps than C, and $T_1$ [mean(C, $T_1$, $T_2$)= 16.8, 17.62, and 10.4; $P_{MW}$($T_2<C$)=0.001; $P_{MW}$($T_2<T_1$)=0.002]. Also, $T_2$ students were observed to derive propositions required to derive $\neg D\lor C$ (for example, $\neg D\lor(C\lor\neg B)$, and $(\neg D\lor C)\lor\neg B$) significantly \textbf{earlier} in their solution attempt than $T_1$ and C [\textbf{For} $\mathbf{\neg D\lor(C\lor\neg B)}$, mean(C, $T_1$, $T_2$)= 22.6, 25.2, and 15.3 mins.; $P_{MW}$($T_2<C$)=0.009; $P_{MW}$($T_2<T_1$)=0.007; \textbf{For} $\mathbf{(\neg D\lor C)\lor\neg B}$, mean(C, $T_1$, $T_2$)= 23.0, 25.6, and 18.08 mins.; $P_{MW}$($T_2<C$)=0.009; $P_{MW}$($T_2<T_1$)=0.005]. Also, $T_2$ had significantly less time gap in between consecutive steps than C [Example: \textbf{for} $\mathbf{\neg D\lor(C\lor\neg B)}$, mean(C, $T_2$)= 7.10, and 3.25 mins.; $P_{MW}$($T_2<C$)=0.012].

On the other hand, the other expert subgoal for A2 is $D\Rightarrow A$ (from region R2). Seven (7) $T_2$ students explicitly derived this proposition backwards. Overall, when adopted approach A2, $T_2$ students discovered subgoal $A\Rightarrow C$ with significantly fewer unnecessary proposition derivations than that of C [Mean unnecessary proposition count before deriving $A\Rightarrow C$  for C, and $T_2$ = 14.90, and 11.25; $P_{MW}$($T_2<C$)=0.001].

These statistics show that $T_2$ students not only identified complex subgoals early in their solution attempts (with less unnecessary propositions, and time), but they were also able to figure out a plan to derive those subgoals (as indicated by early derivation of prerequisite propositions and quicker consecutive steps). Possibly, having the BW skill motivated BW thinking that helped to identify subgoals, and an outline of the solution of the problems, which overall decreased the time required to solve it. However, explicit BW derivations were observed only in 19 out of 54 ($\sim35\%$) $T_2$ students.

\section{Discussion}
In this study, we explored BWE and BPS within an intelligent logic tutoring system to help students adapt to backward strategy use with an aim of improving their subgoaling and problem-solving skills. Our results showed the effectiveness of our training method and revealed important insights on how backward strategy learning impact students' competence in problem-solving. We have summarized the major findings below.

\textbf{\textit{RQ1} (Training Struggle and Increased BW Strategy Usage):} Our results showed that BPS problems caused struggle for students during training (students needed more time, sessions, and restarts). Note that prior studies claim that students usually find backward derivations difficult~\citep{matsuda2005advanced}. Also, experts mostly switch between strategies during problem-solving, unlike BPS problems where students were required to construct the entire solutions backwards, making the training highly complex from a cognitive point of view. However, prior studies claim that challenging, and complex activities increase motivation, induce students to engage deeply, and pay more attention, which helps them to find intrinsic patterns/connections among different parts of a problem, and eventually learn better~\citep{csikszentmihalyi1990flow,bronfenbrenner1979ecology,fullagar2013challenge}. Conforming to this claim, our later findings showed that $T_2$ students voluntarily engaged in backward derivations while solving new problems, and also eventually outperformed the other groups. On the other hand, $T_1$ students receiving an easier training only involving BWEs behaved and performed like the control group.    

\textbf{\textit{RQ2} (Improved Problem Solving Achieved Over Time):} The results From our performance analyses showed that the combination of demonstration (BWE), and practice (BPS) helped students to adapt with backward strategy and improved their problem-solving performance (higher scores, decreased problem-solving time, and step counts). However, improved performance was not observed immediately after students were exposed to the BW strategy. In the earlier phases of training (2.4-4.4), we observed $T_2$ to spend significantly more time while solving simple problems leading to lower scores. As training progressed, $T_2$ increasingly became more efficient in problem-solving and outperformed C, and $T_1$. Recall that BWE/BPSs were given to students mostly during the first half of training [Figure \ref{fig:prob_org}]. However, $T_2$ students continued to improve throughout later phases of training, and posttests. This pattern suggests that BWE+BPS training may require allowing students enough time to adapt to and become adept with the BW strategy, before they can successfully integrate it into their problem-solving approach.

\textbf{\textit{RQ3} (Improved Subgoaling Skill):} Our approach map analyses revealed that $T_2$ students derived expert-identified subgoal propositions more efficiently (with less time, and fewer unnecessary derivations) than the other two groups. On the other hand, $T_1$ did not show any significant consistent evidence of improved subgoaling, possibly due to superficial exposure to BW strategy through BWEs only. Our analyses also showed that, although $T_2$ students engage in BW derivations more than the C, and $T_1$, not all $T_2$ students derived explicit BW steps. However, overall improved subgoaling behavior of $T_2$ students hints at implicit BW strategy use where students formed subgoals using BW thinking, but generated nodes only in the forward direction while working in the DT system, much like experts often do.

\section{Conclusion}
In this paper, we explored backward strategy learning as a way to improve students' subgoaling and problem-solving skills. From our results, we conclude that a combination of examples, active engagement through problem-solving, and enough time allocation is necessary to help and motivate students to learn to use backward strategy. Also, learning backward strategy was proved to help students to derive subgoals and outline solution approaches efficiently (in less time, and less exploration through unnecessary propositions) which led to improved problem-solving performance. As a future work, mixed FW/BW problems where students need to carry out some steps backwards (instead of all backward steps) can be explored for backward strategy learning, that might ensure an appropriate level of cognitive engagement and complexity promoting learning while reducing training struggle.


\backmatter

\bmhead{Supplementary information}

Details of statistical analyses results and Approach map generation procedure can be found in this link:   \href{https://1drv.ms/b/s!AiFP3IlZKTk2cvMDoNQP3Qtg4f4}{https://1drv.ms/b/s!AiFP3IlZKTk2cvMDoNQP3Qtg4f4}.\\\\
Demonstration videos for BWE and BPS can be found in the following links: \\\textbf{BWE} (\href{https://1drv.ms/v/s!AiFP3IlZKTk2cG3tMr0XCm9CnCY}{https://1drv.ms/v/s!AiFP3IlZKTk2cG3tMr0XCm9CnCY}); 
\\\textbf{BPS} (\href{https://1drv.ms/v/s!AiFP3IlZKTk2cWNBVuCfzmaEdsE}{https://1drv.ms/v/s!AiFP3IlZKTk2cWNBVuCfzmaEdsE}).

\bmhead{Acknowledgments}

This material is based upon work supported by the National Science Foundation under Grant No. 2013502.

\bibliography{sn-bibliography}


\section*{Statements and Declarations}

Some journals require declarations to be submitted in a standardised format. Please check the Instructions for Authors of the journal to which you are submitting to see if you need to complete this section. If yes, your manuscript must contain the following sections under the heading `Declarations':

\begin{itemize}
\item \textbf{Funding:} The work was supported by NSF grant 2013502.
\item  \textbf{Conflict of interest/Competing interests:} The authors have no relevant financial or non-financial interests to disclose.
\item  \textbf{Authors' contributions:}
All authors contributed to the study conception and design. Material preparation, data collection and analysis were performed by Preya Shabrina. The first draft of the manuscript was written by Preya Shabrina and all authors commented on previous versions of the manuscript. All authors read and approved the final manuscript.
\end{itemize}

\end{document}

%% file: tables/prelim_stat.tex
\begin{table*}[]
\caption{Training Phase Metrics Values (avg.) across C, $T_1$, and $T_2$(For $T_2$, metric values are shown separately for BPS, and PS). }
\begin{tabular}{|l|l|l|l|l|l|}
\hline
{\color[HTML]{000000} Group} & {\color[HTML]{000000} \begin{tabular}[c]{@{}l@{}}Step Time \\ (min)\end{tabular}} & {\color[HTML]{000000} \begin{tabular}[c]{@{}l@{}}Prob. Time \\ (min)\end{tabular}} & {\color[HTML]{000000} Step Ct.}                                                           & {\color[HTML]{000000} Restarts}                                                     & {\color[HTML]{000000} Sessions}                                                 \\ \hline
{\color[HTML]{000000} C}     & {\color[HTML]{000000} 3.3}                                                        & {\color[HTML]{000000} 36}                                                          & {\color[HTML]{000000} $\sim$10}                                                           & {\color[HTML]{000000} 0.38}                                                         & {\color[HTML]{000000} 4.2}                                                      \\ \hline
{\color[HTML]{000000} T1}    & {\color[HTML]{000000} 1.7}                                                        & {\color[HTML]{000000} 33}                                                          & {\color[HTML]{000000} $\sim$8}                                                            & {\color[HTML]{000000} 0.30}                                                         & {\color[HTML]{000000} 4.1}                                                      \\ \hline
{\color[HTML]{000000} T2}    & {\color[HTML]{000000} \begin{tabular}[c]{@{}l@{}}PS: 2.8\\ BPS: 5.7\end{tabular}} & {\color[HTML]{000000} \begin{tabular}[c]{@{}l@{}}PS: 39\\ BPS: 49.8\end{tabular}}  & {\color[HTML]{000000} \begin{tabular}[c]{@{}l@{}}PS: $\sim$9\\ BPS: $\sim$8\end{tabular}} & {\color[HTML]{000000} \begin{tabular}[c]{@{}l@{}}PS: 0.32\\ BPS: 0.69\end{tabular}} & {\color[HTML]{000000} \begin{tabular}[c]{@{}l@{}}PS: 7.2\\ BPS: 6\end{tabular}} \\ \hline
\end{tabular}
\label{tab:prelim_stat}
\vspace{-2mm}
\end{table*}

%% file: tables/bw_stat.tex
\begin{table*}[]
\caption{Engagement in Backward Actions across the Three Training Groups while Solving Pretest Problems, Training PS Problems (fourth Problem of Level 2-6), and posttest PS Problems. \textbf{Interpreting p-val column:} Read $T_2$ \textgreater $T_1$ (\textless 0.0001) as $T_2$ has significantly higher value than $T_1$ with p<0.0001. $T_2$ \textgreater $T_1$ (0.0001) means p=0.0001 for the hypothesis.}
\begin{tabular}{|l|l|l|l|l|}
\hline
Level (Prob.)  & C    & $T_1$   & $T_2$    & P-val from Mann-Whitney U Tests                                                                                 \\ \hline
Pretest (avg.) & 3.92 & 3.23 & 4.73  & -                                                                                     \\ \hline
2.4          & 2.53 & 8.65 & 58.45 & \textbf{$T_2$\textgreater{}C (\textless 0.0001); $T_2$\textgreater{}$T_1$ (\textless{}0.0001)} \\ \hline
3.4          & 1.63 & 4.13 & 26.35 & \textbf{$T_2$\textgreater{}C (\textless 0.0001); $T_2$\textgreater{}$T_1$ (\textless{}0.0001)} \\ \hline
4.4          & 1.00 & 2.05 & 31.09 & \textbf{$T_2$\textgreater{}C (\textless 0.0001); $T_2$\textgreater{}$T_1$ (\textless{}0.0001)} \\ \hline
5.4          & 0.66 & 1.15 & 3.26  & $T_2$\textgreater{}C (0.015)                                                             \\ \hline
6.4          & 1.29 & 0.87 & 3.31  & \textbf{$T_2$\textgreater{}C (0.0004) ; $T_2$\textgreater{}$T_1$(0.015)}                       \\ \hline
7.1          & 0.22 & 0.29 & 0.57  & -                                                                                     \\ \hline
7.2          & 0.15 & 0.09 & 0.31  & -                                                                                     \\ \hline
7.3          & 0.25 & 0.8  & 3.43  & \textbf{$T_2$\textgreater{}C (0.003) ; $T_2$\textgreater{}$T_1$ (0.02)}          \\ \hline
7.4          & 0.61 & 0.76 & 4.5   & \textbf{$T_2$\textgreater{}C (0.011); $T_2$ \textgreater $T_1$ (0.006)}                        \\ \hline
7.5          & 2.64 & 1.22 & 4.04  & \textbf{$T_2$ \textgreater C (0.010); $T_2$ \textgreater $T_1$ (0.015)}                        \\ \hline
7.6          & 1.75 & 6.29 & 11.17 & \textbf{$T_2$ \textgreater C (0.001)}                                                    \\ \hline
\end{tabular}
\label{table:bw_engagement}
\vspace{-4mm}
\end{table*}

%% file: tables/perf_eval.tex
\begin{table*}[]
\caption{Total Time and Step Count for pretest, training (2(4)-6(4)), and post-test(7(1)-7(6)) PS Problems across the Three Training Groups.}
\begin{tabular}{|l|llll|llll|}
\hline
Prob.   & \multicolumn{4}{l|}{Total Time}                                                                                                                                                                          & \multicolumn{4}{l|}{Step Count}                                                                                                                                                             \\ \hline
        & \multicolumn{1}{l|}{C}     & \multicolumn{1}{l|}{T1}    & \multicolumn{1}{l|}{T2}   & p-val                                                                                                              & \multicolumn{1}{l|}{C}    & \multicolumn{1}{l|}{T1}   & \multicolumn{1}{l|}{T2}   & p-val                                                                                                   \\ \hline
Pretest & \multicolumn{1}{l|}{35.7}  & \multicolumn{1}{l|}{33.0}  & \multicolumn{1}{l|}{32.8} & -                                                                                                                  & \multicolumn{1}{l|}{7.1}  & \multicolumn{1}{l|}{7.3}  & \multicolumn{1}{l|}{7.9}  & -                                                                                                       \\ \hline
2.4     & \multicolumn{1}{l|}{28.5}  & \multicolumn{1}{l|}{30.3}  & \multicolumn{1}{l|}{37.9} & \textbf{\begin{tabular}[c]{@{}l@{}}T2 \textgreater C\\ (0.001); \\ T2 \textgreater{}T1\\ (0.011)\end{tabular}}     & \multicolumn{1}{l|}{7.6}  & \multicolumn{1}{l|}{7.8}  & \multicolumn{1}{l|}{7.7}  & -                                                                                                       \\ \hline
3.4     & \multicolumn{1}{l|}{27.04} & \multicolumn{1}{l|}{24.24} & \multicolumn{1}{l|}{29.5} & \textbf{\begin{tabular}[c]{@{}l@{}}T2 \textgreater T1\\  (0.007)\end{tabular}}                                     & \multicolumn{1}{l|}{10.4} & \multicolumn{1}{l|}{11.3} & \multicolumn{1}{l|}{11.1} & -                                                                                                       \\ \hline
4.4     & \multicolumn{1}{l|}{31.9}  & \multicolumn{1}{l|}{24.82} & \multicolumn{1}{l|}{33.2} & \textbf{\begin{tabular}[c]{@{}l@{}}T2 \textgreater T1\\ (0.003)\end{tabular}}                                      & \multicolumn{1}{l|}{14.2} & \multicolumn{1}{l|}{12.9} & \multicolumn{1}{l|}{14.8} & -                                                                                                       \\ \hline
5.4     & \multicolumn{1}{l|}{31.1}  & \multicolumn{1}{l|}{26.1}  & \multicolumn{1}{l|}{24.0} & -                                                                                                                  & \multicolumn{1}{l|}{11.5} & \multicolumn{1}{l|}{10.2} & \multicolumn{1}{l|}{10.3} & -                                                                                                       \\ \hline
6.4     & \multicolumn{1}{l|}{30.7}  & \multicolumn{1}{l|}{30.1}  & \multicolumn{1}{l|}{31.4} & -                                                                                                                  & \multicolumn{1}{l|}{14.3} & \multicolumn{1}{l|}{14.9} & \multicolumn{1}{l|}{14.9} & -                                                                                                       \\ \hline
7.1     & \multicolumn{1}{l|}{29.5}  & \multicolumn{1}{l|}{27.0}  & \multicolumn{1}{l|}{26.4} & -                                                                                                                  & \multicolumn{1}{l|}{8.5}  & \multicolumn{1}{l|}{8.8}  & \multicolumn{1}{l|}{9.1}  & -                                                                                                       \\ \hline
7.2     & \multicolumn{1}{l|}{30.2}  & \multicolumn{1}{l|}{27.2}  & \multicolumn{1}{l|}{24.6} & -                                                                                                                  & \multicolumn{1}{l|}{5.1}  & \multicolumn{1}{l|}{5.0}  & \multicolumn{1}{l|}{5.1}  & -                                                                                                       \\ \hline
7.3     & \multicolumn{1}{l|}{24.88} & \multicolumn{1}{l|}{25.5}  & \multicolumn{1}{l|}{12.8} & \textbf{\begin{tabular}[c]{@{}l@{}}T2 \textless C\\ (0.02); \\ T2 \textless T1\\ (0.003)\end{tabular}}             & \multicolumn{1}{l|}{10.5} & \multicolumn{1}{l|}{11.0} & \multicolumn{1}{l|}{9.3}  & \textbf{\begin{tabular}[c]{@{}l@{}}T2 \textless C\\ (0.014); \\ T2 \textless T1\\ (0.009)\end{tabular}} \\ \hline
7.4     & \multicolumn{1}{l|}{25.1}  & \multicolumn{1}{l|}{23.7}  & \multicolumn{1}{l|}{15.7} & \textbf{\begin{tabular}[c]{@{}l@{}}T2 \textless C\\ (0.018);\\ T2 \textless T1\\ (0.0006)\end{tabular}}            & \multicolumn{1}{l|}{11.0} & \multicolumn{1}{l|}{11.1} & \multicolumn{1}{l|}{11.2} & -                                                                                                       \\ \hline
7.5     & \multicolumn{1}{l|}{27.36} & \multicolumn{1}{l|}{25.8}  & \multicolumn{1}{l|}{21.7} & \textbf{\begin{tabular}[c]{@{}l@{}}T2 \textless C\\ (0.001);\\ T2 \textless T1\\ (\textless{}0.0001)\end{tabular}} & \multicolumn{1}{l|}{11.0} & \multicolumn{1}{l|}{10.5} & \multicolumn{1}{l|}{9.0}  & \textbf{\begin{tabular}[c]{@{}l@{}}T2 \textless C\\ (0.013);\\ T2 \textless T1\\ (0.012)\end{tabular}}  \\ \hline
7.6     & \multicolumn{1}{l|}{29.56} & \multicolumn{1}{l|}{30.1}  & \multicolumn{1}{l|}{22.7} & \textbf{\begin{tabular}[c]{@{}l@{}}T2 \textless T1\\ (0.010)\end{tabular}}                                         & \multicolumn{1}{l|}{15.1} & \multicolumn{1}{l|}{15.0} & \multicolumn{1}{l|}{14.2} & -                                                                                                       \\ \hline
\end{tabular}
\label{table:perf_eval}
\vspace{-2mm}
\end{table*}

%% file: tables/prob_score.tex
\begin{table*}[]
\caption{Scores for pretest, training (2(4)-6(4)), and post-test(7(1)-7(6)) PS Problems across the Three Training Groups.}
\begin{tabular}{|l|llll|}
\hline
\begin{tabular}[c]{@{}l@{}}Level\\ (Prob.)\end{tabular} & \multicolumn{4}{l|}{Score}                                                                                                                       \\ \hline
                                                        & \multicolumn{1}{l|}{C}    & \multicolumn{1}{l|}{T1}   & \multicolumn{1}{l|}{T2}   & p-val                                                        \\ \hline
Pretest                                                 & \multicolumn{1}{l|}{59.7} & \multicolumn{1}{l|}{60.2} & \multicolumn{1}{l|}{61.2} & -                                                            \\ \hline
2.4                                                     & \multicolumn{1}{l|}{66.5} & \multicolumn{1}{l|}{63.8} & \multicolumn{1}{l|}{52.7} & \textbf{T2 \textless C(0.001); T2 \textless T1(0.012)}       \\ \hline
3.4                                                     & \multicolumn{1}{l|}{61.5} & \multicolumn{1}{l|}{67.3} & \multicolumn{1}{l|}{61.5} & -                                                            \\ \hline
4.4                                                     & \multicolumn{1}{l|}{55.8} & \multicolumn{1}{l|}{65.8} & \multicolumn{1}{l|}{57.2} & \textbf{T2 \textless T1 (0.012)}                             \\ \hline
5.4                                                     & \multicolumn{1}{l|}{67.1} & \multicolumn{1}{l|}{73.7} & \multicolumn{1}{l|}{78.6} & \textbf{T2 \textgreater C(0.002)}                            \\ \hline
6.4                                                     & \multicolumn{1}{l|}{58.3} & \multicolumn{1}{l|}{60.3} & \multicolumn{1}{l|}{59.8} & -                                                            \\ \hline
7.1                                                     & \multicolumn{1}{l|}{57.1} & \multicolumn{1}{l|}{57.6} & \multicolumn{1}{l|}{62.7} & -                                                            \\ \hline
7.2                                                     & \multicolumn{1}{l|}{64.7} & \multicolumn{1}{l|}{64.2} & \multicolumn{1}{l|}{66.5} & -                                                            \\ \hline
7.3                                                     & \multicolumn{1}{l|}{72.2} & \multicolumn{1}{l|}{72.0} & \multicolumn{1}{l|}{81.8} & \textbf{T2 \textgreater C(0.002); T2 \textgreater T1(0.003)} \\ \hline
7.4                                                     & \multicolumn{1}{l|}{71.1} & \multicolumn{1}{l|}{72.1} & \multicolumn{1}{l|}{78.5} & \textbf{T2 \textgreater C(0.009); T2 \textgreater T1(0.012)} \\ \hline
7.5                                                     & \multicolumn{1}{l|}{63.8} & \multicolumn{1}{l|}{66.8} & \multicolumn{1}{l|}{77.0} & \textbf{T2 \textgreater C(0.001); T2 \textgreater T1(0.007)} \\ \hline
7.6                                                     & \multicolumn{1}{l|}{64.2} & \multicolumn{1}{l|}{67.1} & \multicolumn{1}{l|}{74.7} & \textbf{T2 \textgreater C(0.009)}                            \\ \hline
\end{tabular}
\label{table:prob_score}
\end{table*}